\documentclass[aps,prl,twocolumn,superscriptaddress]{revtex4}%
\usepackage{epsfig,dsfont,amssymb,amsmath,amsthm,amsfonts,amsbsy,mathrsfs}
\usepackage{graphicx}
\usepackage{amsmath}
\usepackage{amssymb}%
\begin{document}
\title{Addendum to {\em \lq\lq Single photon logic gates using minimum resources"}}
\author{Qing Lin}
\email{qlin@mail.ustc.edu.cn}
\affiliation{College of Information Science and Engineering, Huaqiao University (Xiamen),
Xiamen 361021, China}
\author{Bing He}
\email{heb@ucalgary.ca}
\affiliation{Institute for Quantum Information Science, University of Calgary, Alberta T2N
1N4, Canada}


\begin{abstract}
The authors call attention to a previous work [Qing Lin and Bing He, Phys. Rev. A 80, 042310 (2009)] on the realization of multi-qubit logic gates with controlled-path and merging gate. We supplement the work by showing how to efficiently build realistic quantum circuits in this approach and giving the guiding rules for the task.
\end{abstract}
\maketitle

\begin{figure}[ptb]
\includegraphics[width=8.7cm]{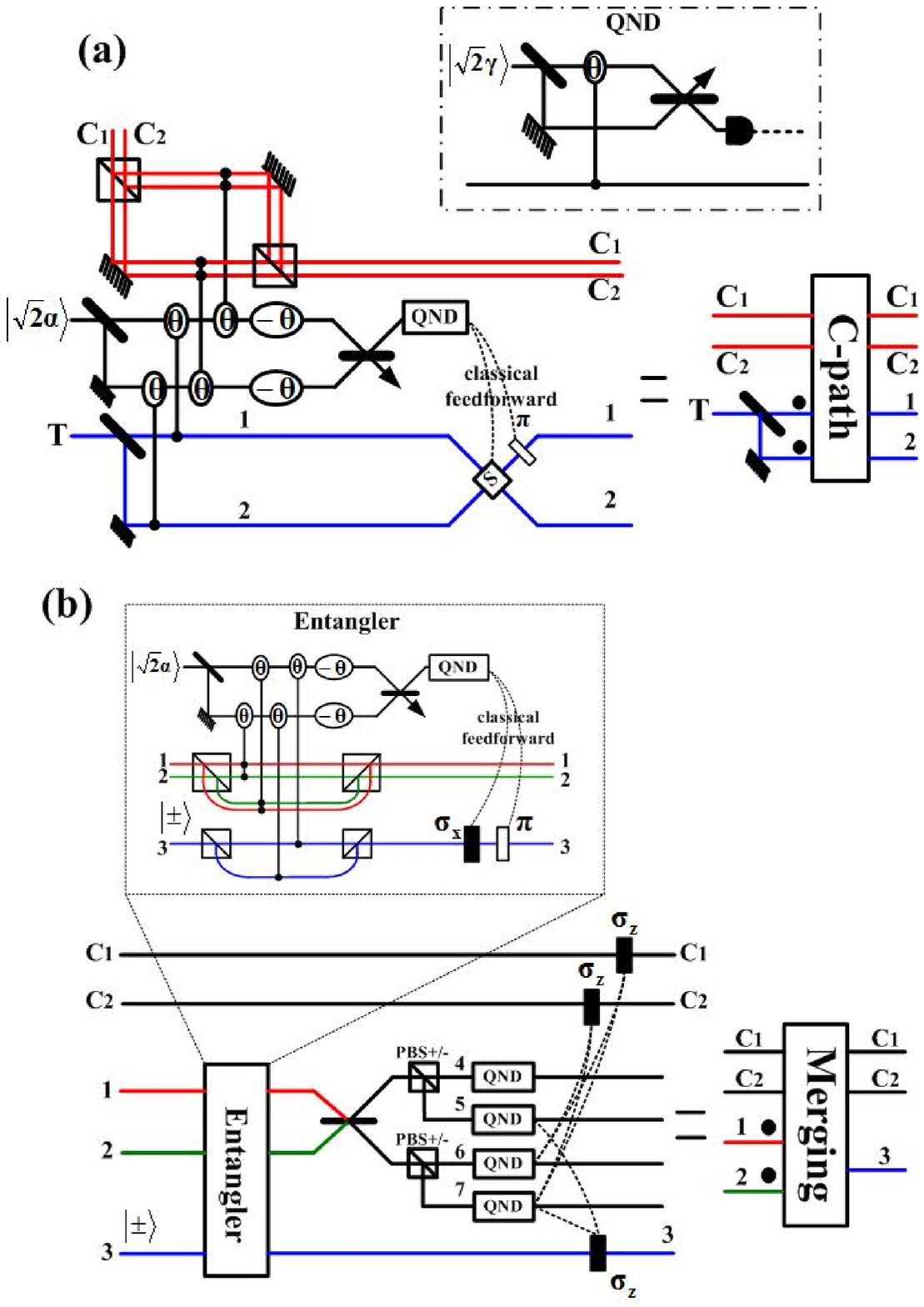}
\caption{(color online) (a) Layout for a modified C-path gate. The control photon
has two different spatial modes
(denoted by black dot), which will interact with two other spatial
modes of the target photon through Kerr media. The XPM phases are $\theta$, and a phase shift $-\theta$ is applied to the two coherent states, respectively.
The part in the dash-dotted line shows the structure of the quantum-nondemolition measurement (QND) module. $S$ denotes a switch operation, and $\pi$ a phase shift of such value. $|\sqrt{2}\alpha\rangle$ and $|\sqrt{2}\gamma\rangle$ are coherent states. (b) Layout of the corresponding merging gate, which implements the inverse operation for the C-path gate.}%
\vspace{-0.5cm}
\end{figure}

A core element for quantum computer is quantum logic gate. The
realization of photonic logic gates is one of the main directions in
the research of optical quantum computing. Compared with the linear
optical approach \cite{klm, l-review}, deterministic photonic gates
enjoy the advantages of efficiency and simplicity. The realizations
of such gates are proposed mostly with nonlinear optical process
such as cross phase modulation (XPM) in Kerr media \cite{p-p}. In
this approach, qubits in the basis $\{|0\rangle,|1\rangle\}$ can be
simply encoded with photon number, polarization or spatial modes of
single photons, and a gate operation dispenses with complicated
ancilla photonic states. For example, an XPM between two photons
directly implements a controlled phase gate with the transformation
$|1\rangle_1\otimes|1\rangle_2\rightarrow
e^{i\theta}|1\rangle_1\otimes|1\rangle_2$ (with
$|0\rangle_1\otimes|0\rangle_2$, $|0\rangle_1\otimes|1\rangle_2$ and
$|1\rangle_1\otimes|0\rangle_2$ being unchanged). However, due to
technical challenges, a value of $\theta$ in the order of $\pi$
radian could be difficult to come by. One feasible substitute is the
weak nonlinearity approach \cite{w-n1,w-n2}. It is to couple
coherent states $|\alpha\rangle$ of a large amplitude $|\alpha|$ to
single photons for the transformation
$|\alpha\rangle_1|1\rangle_2\rightarrow |\alpha
e^{i\theta}\rangle_1|1\rangle_2$, where $\theta$ could be rather
small. Then the processed coherent states are measured for
projecting out the proper output states of the single photons, which
should be obtained by gate operation. Recently we developed the
approach further by proposing an architecture using two element
gates---controlled-path (C-path) and merging gate---for
single-photon logic gates  \cite{Lin2}. Multiple qubit control gates
such as the Fredkin and the Toffoli gate, which are under extensive studies
recently \cite{a1,a2,a3,a4,a5,a6,a7,a8}, can be efficiently realized by the combinations of the two
element gates. This Brief Report supplements the previous study from
the view of constructing the realistic circuits for implementing
quantum algorithms.

A quantum circuit consists of various ingredients, e.g., single
qubit gates, two-qubit gates, multi-control gates, etc. Individually
all control gates involving more than two qubits can be realized
with pair(s) of C-path and merging gate, together with the necessary
single qubit gates \cite{Lin2}. A prominent feature in a realistic
quantum circuit is that more than one control operation could be
acted on a particular target qubit, so the photon to encode the
target qubit should be separated by C-path gates and merged by
merging gates again and again if we straightforwardly apply the
elementary gates to implement a circuit operation. Actually such
repetition can be saved if one modifies C-path and merging gate a
little bit. Then a target photon could be merged only after all
control operations have been performed on it, thus greatly
simplifying circuit structure by reducing the number of the merging
gates. Such simplification is particularly relevant to circuit
operations involving large number of qubits, e.g., the
implementation of quantum algorithms. To fulfill the simplification,
a new element---the eraser for eliminating the unwanted photon
correlations between the successive C-path gate operations, should
be introduced, as we will explain below.

The control photon in an original C-path or merging gate
carries only one spatial mode \cite{Lin1, Lin2}. Here we make a modification in the design so that a photon with more than one spatial mode could control the path of another photon (a similar modification in a special case is given in \cite{Lin3}). In Fig. 1(a),
we suppose that the input state (resulting from the action of the previous
logic gates) for the gate is
\begin{eqnarray}
\left\vert \psi\right\rangle _{CT}^2&=&\left\vert H\right\rangle _{C_{1}}\left\vert \phi_{1}\right\rangle
_{T}+\left\vert H\right\rangle _{C_{2}}\left\vert \phi_{2}\right\rangle
_{T}+\left\vert V\right\rangle _{C_{1}}\left\vert \phi_{3}\right\rangle
_{T}\nonumber\\
&+&\left\vert V\right\rangle _{C_{2}}\left\vert \phi_{4}\right\rangle _{T},
\end{eqnarray}
where $C_{1},C_{2}$ denote the different spatial modes of the control photon, $H, V$ denote the polarization modes, and the components of the target photon is $\left\vert \phi
_{i}\right\rangle =\alpha_{i}\left\vert H\right\rangle +\beta_{i}\left\vert
V\right\rangle $, where $\underset{i=1}{\overset{4}{\sum}}\left(  \left\vert
\alpha_{i}\right\vert ^{2}+\left\vert \beta_{i}\right\vert ^{2}\right) =1$.
The special forms of such inputs with $\alpha_{1,3}=\beta_{1,3}=0$ or
$\alpha_{2,4}=\beta_{2,4}=0$ can be processed by the original C-path gate in
\cite{Lin2, Lin3, Lin1}. We first use a 50:50 beam splitter (BS) to divide the target photon into two different spatial modes:
\begin{align}
&  \frac{1}{\sqrt{2}}\left[  \left\vert H\right\rangle _{C_{1}}\left(
\left\vert \phi_{1}\right\rangle _{1}+\left\vert \phi_{1}\right\rangle
_{2}\right)  +\left\vert H\right\rangle _{C_{2}}\left(  \left\vert \phi
_{2}\right\rangle _{1}+\left\vert \phi_{2}\right\rangle _{2}\right)  \right.
\nonumber\\
&  \left.  +\left\vert V\right\rangle _{C_{1}}\left(  \left\vert \phi
_{3}\right\rangle _{1}+\left\vert \phi_{3}\right\rangle _{2}\right)
+\left\vert V\right\rangle _{C_{2}}\left(  \left\vert \phi_{4}\right\rangle
_{1}+\left\vert \phi_{4}\right\rangle _{2}\right)  \right]  ,
\label{c-path}
\end{align}
where the index $1$ and $2$ denote two different paths. And then, following the coupling patterns of XPM in Fig. 1(a), i.e., the first (second) coherent state $|\alpha\rangle$ is coupled to mode 1 (2) of the target and $V$ ($H$) mode on both $C_1$ and $C_2$ for the control photon, we will obtain the following total state (the global coefficient is neglected):
\begin{eqnarray}
&& \left\vert H\right\rangle _{C_{1}}\left(
\left\vert \phi_{1}\right\rangle _{1}|\alpha e^{i\theta}\rangle |\alpha e^{i\theta}\rangle+\left\vert \phi_{1}\right\rangle
_{2}|\alpha\rangle|\alpha e^{2i\theta}\rangle\right)\nonumber\\
&+&\left\vert H\right\rangle _{C_{2}}\left(  \left\vert \phi
_{2}\right\rangle _{1}|\alpha e^{i\theta}\rangle|\alpha e^{i\theta}\rangle+\left\vert \phi_{2}\right\rangle _{2}|\alpha\rangle|\alpha e^{2i\theta}\rangle\right)\nonumber\\
&+&\left\vert V\right\rangle _{C_{1}}\left(  \left\vert \phi
_{3}\right\rangle _{1}|\alpha e^{2i\theta}\rangle|\alpha\rangle+\left\vert \phi_{3}\right\rangle _{2}|\alpha e^{i\theta}\rangle|\alpha e^{i\theta}\rangle\right)\nonumber\\
&+&\left\vert V\right\rangle _{C_{2}}\left(  \left\vert \phi_{4}\right\rangle
_{1}|\alpha e^{2i\theta}\rangle|\alpha\rangle+\left\vert \phi_{4}\right\rangle _{2}|\alpha e^{i\theta}\rangle|\alpha e^{i\theta}\rangle\right).
\end{eqnarray}
After a phase shifter $-\theta$ and a 50:50 beam splitter (BS) implementing the transformation $\left\vert
\alpha_{1}\right\rangle \left\vert \alpha_{2}\right\rangle \rightarrow
\left\vert \frac{\alpha_{1}-\alpha_{2}}{\sqrt{2}}\right\rangle \left\vert
\frac{\alpha_{1}+\alpha_{2}}{\sqrt{2}}\right\rangle $ are performed on the two ancilla beams,
the eight terms in the above equation can be projected into two groups of output state by
a photon number projection $|n\rangle\langle n|$ on one of the output beams, which could be in the state $|\beta\rangle=|\pm \sqrt{2}\alpha\sin\theta\rangle$ or $|0\rangle$. By the path switch $S$ and a phase shift $\pi$ on the target photon, which is conditioned on the measurement results $n\neq 0$, the two-photon state from both of the groups can be transformed to
\begin{equation}
\left\vert \phi\right\rangle =\left\vert H\right\rangle _{C_{1}}\left\vert
\phi_{1}\right\rangle _{1}+\left\vert H\right\rangle _{C_{2}}\left\vert
\phi_{2}\right\rangle _{1}+\left\vert V\right\rangle _{C_{1}}\left\vert
\phi_{3}\right\rangle _{2}+\left\vert V\right\rangle _{C_{2}}\left\vert
\phi_{4}\right\rangle _{2},
\label{path}
\end{equation}
thus realizing a deterministic control of the target photon's paths
by the polarizations ($H$ and $V$) of the control photon carrying
two spatial modes. The projection $|n\rangle\langle n|$ is
implemented by a QND module, in which a beam in the state
$|\gamma\rangle$ (where $|\gamma|$ is large) is coupled to the
above-mentioned output ancilla beam through an XPM implementing the
transformation $|\beta\rangle|\gamma\rangle\rightarrow
e^{-\frac{1}{2}|\beta|^2}(|0\rangle
|\gamma\rangle+\beta|1\rangle|\gamma e^{i\theta}\rangle+\cdots)$.
Even if $\theta\ll 1$, the output coherent states $|\frac{\gamma
e^{in\theta}\pm \gamma }{\sqrt{2}}\rangle$ can be still well
separated with respect to their Poisson distributions of photon
numbers, given a sufficiently large $|\gamma|$. Thus, a number
non-resolving detector even without high detecting efficiency or a
quadrature measurement can indirectly realize the deterministic
photon number resolving detection corresponding to $|n\rangle\langle
n|$ (see \cite{Lin3} for the details). The progress on the physical
realization of such XPM based QND refers to, e.g.,
\cite{qnd1,qnd2,qnd3,qnd4}. The number of the controlling spatial
modes for the C-path gate can be straightforwardly generalized to
larger than two.

Similarly, we can modify a merging gate, which performs the inverse operation of the above C-path gate, see Fig. 1(b). By such merging gate with multi-spatial control modes, the output state in Eq. (\ref{path}) can be transformed to
\begin{equation}
\left\vert H\right\rangle _{C_{1}}\left\vert \phi_{1}\right\rangle
_{3}+\left\vert H\right\rangle _{C_{2}}\left\vert \phi_{2}\right\rangle
_{3}+\left\vert V\right\rangle _{C_{1}}\left\vert \phi_{3}\right\rangle
_{3}+\left\vert V\right\rangle _{C_{2}}\left\vert \phi_{4}\right\rangle _{3},
\label{merge}
\end{equation}
i.e., the merging of the target photon modes on path 1 and 2 to path 3.

Now we will put the element gates together to build a quantum circuit.
Without loss of generality, we illustrate the architecture by a three-qubit circuit shown
in dash-dotted line of Fig. 2. The input state for the circuit is (the global coefficient is neglected)
\begin{equation}
|\psi_{in}\rangle=d_1|HHH\rangle+d_2|HHV\rangle+\cdots+d_8|VVV\rangle,
\end{equation}
which could be either entangled or not entangled.
Three CU operations and one Toffoli operation will be performed on the state.
In the space where the qubits are encoded with the polarization modes of single photons, the CU operations are represented by the operators $|H\rangle\langle H|\otimes \mathbb{I}+|V\rangle\langle V|\otimes U_i $, where $\mathbb{I}=|H\rangle\langle H|+|V\rangle\langle V|$ and $i=1,2,4$. The Toffoli gate performs the operation
$(\mathbb{I}\otimes \mathbb{I}-|VV\rangle\langle VV|)\otimes \mathbb{I}+|VV\rangle\langle VV|\otimes U_3 $.

\begin{figure}[b!]
\includegraphics[width=9cm]{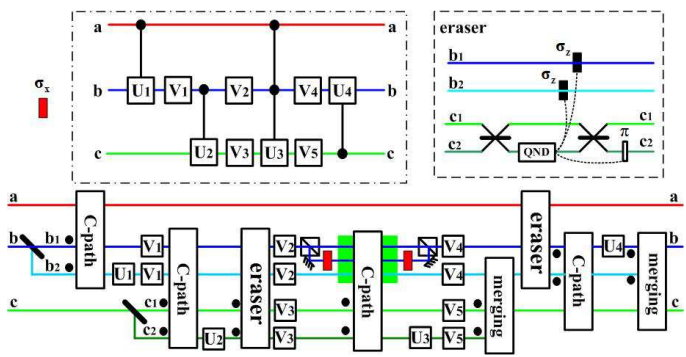}
\caption{(color online) Example of triple-qubit gate. The circuit in the dash-dotted line is implemented by the modified C-path and merging gates. A necessary ingredient for the implementation is the eraser shown in the dashed line. }%
\end{figure}

The first CU gate can be straightforwardly realized with a C-path gate plus single qubit operation $U_1$.
Before the implementation of the second CU operation involving photon $b$ and $c$, we do
not merge the spatial modes of photon $b$ and, instead, we directly use them to control the operation on the third
photon in the next gate operation. The second CU operation can be implemented by a
generalized C-path gate in Fig. 1(a), associated with the single-photon operation $U_{2}$
performed on the spatial mode $c_{2}$.

The triple photon state after being processed by the first two CU gates is in the form
\begin{eqnarray}
&&\left\vert H\right\rangle _{a}\left\vert H\right\rangle _{b_{1}}\left\vert \phi_{1}\right\rangle _{c_{1}%
}+\left\vert V\right\rangle _{a}\left\vert H\right\rangle _{b_{2}}\left\vert \phi_{2}\right\rangle _{c_{1}%
}\nonumber\\
&+&\left\vert H\right\rangle _{a}\left\vert V\right\rangle _{b_{1}}\left\vert \phi_{3}\right\rangle _{c_{2}%
}
+\left\vert V\right\rangle _{a}\left\vert V\right\rangle _{b_{2}}\left\vert \phi_{4}\right\rangle _{c_{2}},
\label{in}
\end{eqnarray}
where the specific forms of $\left\vert \phi_{i}\right\rangle _{c_{j}}$ are determined by the operations $U_i$, $V_i$ and the coefficients $d_i$ of the input state.
The polarization modes of photon $b$ are entangled with the spatial modes of photon $c$ in the above expression ($|H\rangle$ of photon $b$ is always in the same terms with the spatial mode $1$ of photon $c$, etc.). However, a proper state to be processed by the next C-path gate should
be in the form of Eq. (\ref{c-path}), where each polarization mode (irrespective of its spatial mode) for the control photon is in tensor product with the superposition of both spatial mode terms of the target photon. It is therefore necessary to eliminate such unwanted correlation between the polarizations of photon $b$ and spatial modes of photon $c$.

We introduce a circuit ingredient illustrated in the dashed
line of Fig. 2 for the purpose. The QND module here is the same as the previously described.
After the interference of the spatial modes $c_1$ and $c_2$ by a 50:50 beam splitter (BS), a QND module projects the state to
\begin{align}
&  \left\vert HH\right\rangle _{a, b_{1}}\left\vert \phi_{1}\right\rangle _{c_{1}%
}+\left\vert VH\right\rangle _{a,b_{2}}\left\vert \phi_{2}\right\rangle _{c_{1}%
}+\left\vert HV\right\rangle _{a,b_{1}}\left\vert \phi_{3}\right\rangle _{c_{1}}\text{ }\nonumber\\
&+\left\vert VV\right\rangle _{a,b_{2}}\left\vert \phi_{4}\right\rangle _{c_{1}%
}\text{ }\nonumber\\
&  or\text{ }  \left\vert HH\right\rangle _{a, b_{1}}\left\vert \phi_{1}\right\rangle _{c_{2}%
}+\left\vert VH\right\rangle _{a,b_{2}}\left\vert \phi_{2}\right\rangle _{c_{2}%
}-\left\vert HV\right\rangle _{a,b_{1}}\left\vert \phi_{3}\right\rangle _{c_{2}}\text{ }\nonumber\\
&-\left\vert VV\right\rangle _{a,b_{2}}\left\vert \phi_{4}\right\rangle _{c_{2}%
}.
\end{align}
Then one more BS, as well as the conditional operation $\sigma_{z}$ on both spatial modes of photon $b$
and a conditional phase shift $\pi$ on the mode $c_{2}$,
depending on the classically feedforwarded detection results of the QND, will be applied. The whole operation will result in the state
\begin{align}
&  \frac{1}{\sqrt{2}}\left[  \left\vert HH\right\rangle _{a,b_{1}}\left(
\left\vert \phi_{1}\right\rangle _{c_{1}}+\left\vert \phi_{1}\right\rangle
_{c_{2}}\right)  +\left\vert VH\right\rangle _{a,b_{2}}\left(  \left\vert
\phi_{2}\right\rangle _{c_{1}}+\left\vert \phi_{2}\right\rangle _{c_{2}%
}\right)  \right.  \nonumber\\
&  +\left.  \left\vert HV\right\rangle _{a,b_{1}}\left(  \left\vert \phi
_{3}\right\rangle _{c_{1}}+\left\vert \phi_{3}\right\rangle _{c_{2}}\right)
+\left\vert VV\right\rangle _{a,b_{2}}\left(  \left\vert \phi_{4}\right\rangle
_{c_{1}}+\left\vert \phi_{4}\right\rangle _{c_{2}}\right)  \right]  ,
\label{term}
\end{align}
which is similar to the form of the input in Eq. (2). We call this circuit ingredient an eraser.

Now the correlation between the polarization and spatial modes of
photon $a$ and $b$ still exists, i.e., the mode $1$ of photon $b$ is
always in the same terms with $H$ mode of photon $a$,  etc., see Eq.
(\ref{term}). This happens to be an advantage for implementing the
following Toffoli gate. Then a C-path gate only by photon $b$'s
polarizations on its all spatial paths will control the path of
photon $c$, thus realizing a Toffoli gate with a simplified
structure from that in \cite{Lin2}.

It is not necessary to erase the correlation between the photons immediately after implementing the Toffoli gate, but the spatial modes of the third photon should be merged by a merging gate. There will be no control operation to be performed on this photon (photon $c$). The merging of its spatial modes will simplify the further control operation by itself. Before the final CU operation, one more eraser should be applied to eliminate the unwanted correlation between photon $a$ and $b$.

The imperfections of the circuit operation could arise from losses
of the photonic states. The decoherence effects on photonic states
due to the losses in XPM and transmission are studied with the
corresponding master equations in \cite{deco1,deco2,deco3}. In the
implementation of the above circuit, the losses will mainly come
from the XPM between single photon and coherent state if there is
close to ideal performance of the linear optical components. It is
shown in \cite{deco1} that, an XPM in lengthy optical fiber is
impossible to avoid considerable decoherence, but the acceptable
fidelities (with the ideal pure output states) could be achievable
with the reasonable system parameters for the XPM in media under
electromagnetically induced transparency (EIT) conditions. A feasible
realization of the XPM between coherent and single photon state still
awaits to be clarified further for building such photonic circuits.

\begin{figure}[t!]
\includegraphics[width=7.2cm]{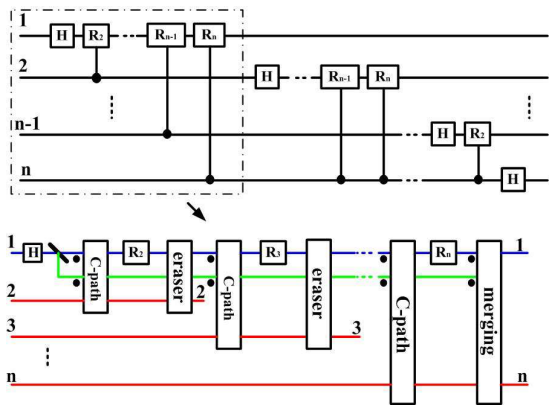}
\caption{(color online) Upper: quantum Fourier transformation
(QFT). The QFT circuit consists of a series of controlled rotations
\cite{Nielsen}. Lower: the realization of the part in the dash-dotted line with the modified C-path, merging gates, together with erasers and single-qubit gates.  }%
\vspace{-0.6cm}
\end{figure}

The generalization to the circuits involving more qubits is
straightforward. Here we give two examples for the implementation of quantum algorithms.
The first is the Grover's searching algorithm \cite{grover}, which could be implemented with $O(n^{2})$ two-qubit gates \cite{Toffoli}. The essential part of the algorithm is the following operation
\begin{equation}
U_{s}=2\left\vert s\right\rangle \left\langle s\right\vert -\mathbb{I}=H^{\otimes n}\left(  2\left\vert 0\right\rangle \left\langle
0\right\vert -\mathbb{I}\right)  H^{\otimes n},
\end{equation}
where $\mathbb{I}$ is the identity operator for $n$ qubits, and $\left\vert
s\right\rangle =H^{\otimes n}\left\vert 0\right\rangle $
($H^{\otimes n}$ denotes $n$ Hadamard operations on $n$ qubits, respectively).
The operator $2\left\vert 0\right\rangle \left\langle 0\right\vert
-\mathbb{I}$ in the above equation is an $n-1$-control Toffoli-$\sigma_{z}$ gate, i.e., the
logic $1$ of the first $n-1$ qubits conditions the operation $\sigma_{z}$ on the $n$-th qubit. Therefore, as a simple generalization from the two-qubit Toffoli gate in Fig. 2, it can be implemented with $n-1$ pairs of C-path and merging gates. Since the paths of the $k$-th photon to encode the qubit is only controlled by the polarization modes of the $k-1$-th photon, there are at most $5+7(n-2)$ XPM operations including those in the QND modules for all C-path gates. Together with those of the merging gates, the total number of XPM operations should be $18n-20$, scaling linearly with the number of the involved qubits. Note that erasers are not necessary in this case. With the QND modules in the merging gates for preserving ancilla photon, in principle only one ancilla photon will be required for implementing the searching algorithm.

Another example is the Shor's factoring algorithm \cite{shor}.
Quantum Fourier transformation (QFT) shown in the upper side of Fig. 3 is the crucial part for the
algorithm. A QFT circuit consists of a series of qubit rotations $R_i$ controlled by other qubits. It can be realized with a regular combination of C-path and merging gates, see the lower side of Fig. 3. A general QFT circuit involving $n$ qubits consists of $n(n-1)/2$ controlled rotations. It therefore demands $n(n-1)/2$ C-path gates, $(n-1)(n-2)/2$ erasers and $n-1$ merging gates for the realization in our architecture. Meanwhile, two points should be paid attention to in constructing such circuit:

(1) before implementing a two-qubit control gate, the correlation between the target photon and the previous control photon should be erased;

(2) if no further control operation is to be performed on a photon,
its spatial modes should be merged with a merging gate.

Following these two rules, it is convenient to construct any quantum circuit with the modified C-path and merging gates. If one adopts the routine decomposition strategy into two-qubit and single-qubit gates for the realization of a quantum circuit (see, e.g., \cite{Nielsen}), the decomposition is generally irregular and the design could be rather complicated. Our architecture following the simple rules of combining C-path, merging gates, as well as erasers, considerably reduces such complexity.

\begin{acknowledgments}
The authors thank Ru-Bing Yang for helpful
suggestions. Q. L. was funded by National Natural Science Foundation
of China (Grant No. 11005040) and the Natural Science Foundation of
FuJian Province of China (Grant No. 2010J05008), and B. H.
acknowledges the support by Alberta Innovates.
\end{acknowledgments}

\end{document}